\documentclass{aa}

\usepackage{amsmath,amssymb}

\newcounter{subequation}
\renewcommand{\theequation}{\arabic{equation}\ifnum\thesubequation>0{\alph{subequation}}\fi}
\newcommand{\subnumbers}{\setcounter{subequation}{1}}
\newcommand{\nosubnumbers}{\setcounter{subequation}{0}}
\newcommand{\stepsubnumber}{\addtocounter{subequation}{1}
\addtocounter{equation}{-1}}
\nosubnumbers

\begin{document}

\thesaurus{02.07.1; 02.08.1; 02.09.1; 03.13.1; 12.03.4; 12.12.1}

\title{Modeling multi--stream flow in collisionless matter:
approximations for large--scale structure\\ beyond shell--crossing}
\author{Thomas Buchert\inst{1} \and Alvaro Dom\'\i nguez\inst{2}} 
\offprints{T. Buchert}
\institute{Theoretische Physik, Ludwig--Maximilians--Universit\"at, 
Theresienstr. 37, D--80333 M\"unchen, Germany
\and
Laboratorio de Astrof\'\i sica Espacial y F\'\i sica Fundamental,
Apartado 50727, E--28080 Madrid, Spain}

\date{Received 29. September 1997, accepted 2. March 1998}

\titlerunning{Modeling multi--stream flow in collisionless matter}

\maketitle

\begin{abstract}

The generally held view that a model of large--scale structure, 
formed by collisionless matter in the Universe, can be based on
the matter model ``dust'' fails in the presence of multi--stream
flow, i.e., velocity dispersion. We argue that models for large--scale
structure should rather be constructed for a flow which describes the
average motion of a multi--stream system.  We present a clearcut
reasoning how to approach the problem and derive an evolution equation for
the mean peculiar--velocity relative to background
solutions of Friedmann--Lema\^\i tre type.  We consider restrictions
of the nonlinear problem and show that the effect of velocity
dispersion gives rise to an effective viscosity of non--dissipative
gravitational origin. We discuss subcases which arise naturally
from this approach: the ``sticky particle model'' and the ``adhesion
approximation''. We also construct a novel approximation that
features adhesive action in the multi--stream regime while conserving
momentum, which was considered a drawback of the standard
approximation based on Burgers' equation. We finally argue that the
assumptions made to obtain these models should be relaxed and we
discuss how this can be achieved.


\end{abstract}

\section{Why model a multi--stream flow ?}

The problem of how to treat a multi--stream flow and the question of 
how to follow approximations beyond shell--crossing time are themes that are 
often repeated in the history of constructing models for large--scale structure. The
application of approximation schemes such as, e.g., Lagrangian
perturbation solutions (which, to first order, contain the celebrated
``Zel'dovich approximation'' (Zel'dovich 1970, 1973) as a special
case, see Buchert 1989, 1992) breaks down at the epoch of formation
of caustics in the density field.  Fluid elements are treated as to
propagate freely through each other; multi--stream flow arises beyond
the caustic, i.e., the velocity field is no longer
single--valued. Self--gravitation of the multi--stream system is the
source for holding structures together: a considerable fraction of the
particles is trapped within three--stream systems (called `pancakes'
in the cosmological context) which, in the course of time, develop a
hierarchy of nested structures due to the formation of N--stream
systems with increasing N. Doroshkevich et al. (1980) have
demonstrated this in a two--dimensional numerical simulation;
the number of streams as a function of time was calculated by Kofman 
et al. (1994).  
This hierarchy of structures is also predicted 
by higher--order Lagrangian perturbation
schemes (Buchert et al. 1997); compare also the works by 
Shukurov (1981) and Fillmore \& Goldreich (1984a,b). The
notion of `non--dissipative gravitational turbulence' has been advanced
for this phenomenon by Gurevich \& Zybin (1995 and ref. therein). For
further discussions in the framework of Lagrangian perturbation theory
on details of multiple shell--crossings related to bifurcations of the
Lagrangian solutions and the mathematical problems involved see, e.g.,
Buchert \& Ehlers (1993), Buchert (1994, 1996),
Ehlers \& Buchert (1997); in the relativistic context: Clarke \&
O'Donnell (1992).

In order to repair this shortcoming, the ``adhesion approximation''
(Gurbatov et al. 1989) has been proposed. This model proved to be
successful, provided the structure formation process is not followed
too deeply into the nonlinear regime (Kofman et al. 1990, 
Weinberg \& Gunn 1990a,b; Sahni \& Coles 1995, Shandarin 1997). At late
nonlinear stages the model predicts a distribution of large--scale
structures which is different from that predicted by N--body
simulations, especially for much small--scale power.
One reason for this was referred to the model itself, i.e. that  
the gravitational potential may acquire an additional
part via nonlinear evolution on smaller scales (Kofman et al. 1992),
it may also be traced back to the model's drawback of not conserving the
linear momentum (Kofman \& Shandarin 1988). We advocate the explanation 
that it may also be a consequence of treating the ``viscosity coefficient'' 
as spatially constant, as will become clear later. 
A recent work by Shandarin \&
Sathyaprakash (1996) proposes a semi--numerical model which is built in the
spirit of `adhesion' of particles, but which has the advantage of
conserving momentum by construction.  
In this paper we will construct models that
allow following the structure formation process beyond shell--crossing
time analytically having both the `adhesive' property of `sticking'
fluid elements and the conservation of momentum.

\noindent
We shall emphasize that the notion of a `self--gravitating gas' (which
we are going to specify) is more adequate when we speak about the
regime of multi--stream flow in collisionless systems than the
generally adopted simplification to ``dust matter'': in the
presence of multi-stream flow, one should focus on the average bulk
motion rather than on individual particles. We shall argue that the
action of the multi--stream system on the average motion can be
modeled by means of pressure--like forces.
Taking this point of view we will be able to propose an approximation
scheme that is valid in the regime of multi--stream flow.  The
assumption of isotropic velocity dispersion singles out a model which
determines a transport coefficient -- not unlike the ``viscosity'' in
the standard ``adhesion approximation'' -- which, however, is of {\it
gravitational origin}. The advantage of this point of view is
two--fold: firstly, we do not have to invent a ``viscosity term'' to
mimic gravitational adhesion of fluid elements after shell--crossing,
but we can {\it derive} it from the gravitational action of the
multi--stream system itself and, secondly, we can learn about the
restrictions we have to impose on the general problem in order to
obtain models of this type.  The latter will strongly support our
opinion that these restrictions should be relaxed.  We wish to note
that, starting from another reasoning, a similar interpretation of the
``viscosity'' in Burgers' equation has been advocated by Ron Kates
(priv. comm. and unpublished notes).

We also think of several implications of such a description,
although we are not going to develop them in this paper. It is known
that ``dust'' is only recovered as a singular limit: a pure
single--stream system maintains a $\delta$--shaped velocity
distribution along trajectories and generates a singularity at
shell--crossing. An initially small deviation from a
$\delta$--distribution allows, however, to follow the distribution
function smoothly beyond shell--crossing; caustics occurring in the
density field for ``dust matter'' are smoothed out by velocity
dispersion (Zel'dovich \& Shandarin 1982). Thus, besides the obvious
advantages implied by a model which can be followed through
shell--crossing, it also entails the description of how velocity
dispersion evolves, which is a key element in the diagnostics of cosmic
structure. The ``Jeans' length'', which has not to be as small as the
one associated with the usual thermostatic pressure, represents the
scale on which structures are diffused (for a discussion of the Jeans'
length in collisionless systems see Seliverstov 1974, for a simulation
of neutrino distributions see Melott 1982,1983, Ma \& Bertschinger
1994), for a Fermi gas a lower limit is set by the phase space
constraint (Tremaine
\& Gunn 1979, Ruffini \& Song 1987, Kull et al. 1997). 
Including velocity dispersion
will allow us to access smaller spatial scales than with previous
analytical models, and gives meaning to the notion of (virial)
equilibrium in a collisionless system, which is a well--studied case in
the context of stellar systems (Binney \& Tremaine 1987). The
treatment of multi--stream flow on cosmological scales will thus
connect models for large--scale structure with small--scale systems
which, apart from recent efforts to simulate galaxy formation embedded
into large--scale structure, have been treated as isolated systems so
far. Yet another perspective is opened concerning reconstruction
models of the density field from observed peculiar--velocity data
(see Dekel 1994 for a review).  Present reconstruction techniques
suffer from the fact that they employ nonlinear models in regimes when
singularities in the flow developed; the quest for singularity--free
models is especially obvious in that context (for a discussion of this
issue see Susperregi \& Buchert 1997).

We proceed as follows: in Sect. 2 
we introduce the system of equations that we are going to study, 
in Sect. 3 we give an equation governing the evolution of the
mean peculiar--velocity in the presence of multi--stream flow, discuss
some properties of the dynamical evolution prescribed by this
equation, and relate it to commonly employed models of large--scale
structure. In Sect. 4 we present our criticism and perspectives.

\section{Basic equations in the presence of velocity dispersion.}  

In this section we shall first use physical (i.e., non--rotating Eulerian)
coordinates. The introduction of so--called `comoving coordinates'
(i.e., scaled Eulerian coordinates indexing fundamental observers 
in a background cosmology) will be employed when 
comparing with commonly used approximation schemes in cosmology in the 
next section.

Let us consider a gas of $N$ particles of mass $m$ which interact only
gravitationally. We may not only think of stars (which is the common
point of view -- see the textbooks by, e.g., Binney \& Tremaine 1987, 
Ogorodnikov 1965 and Saslaw 1985 --), but alternatively 
of dark matter particles (e.g., non--relativistic massive neutrinos or WIMP's, 
e.g., Bertschinger 1993), 
or corresponding larger mass units (like galaxy halos) or, generally, 
of any patch of collisionless matter 
(a fluid element or a Lagrangian particle). The interpretation of our
model depends on the context to which we apply it. In any case we have 
to specify the nature of the particles to see whether the assumptions apply
-- at least in a phenomenological sense -- (e.g., 
for a gas of galaxy halos we neglect merger events, 
or dynamical friction; this also violates
the particle number conservation; for any patch of matter we implicitly 
neglect additional terms that would arise by coarse--graining the 
distribution function, etc.). 

\noindent
We define the one--particle distribution function
$f({\bf x}, {\bf v}, t)$ such that $f({\bf x}, {\bf v}, t) d{\bf x}
d{\bf v}$ is the probability density that a particle be within the volume
element $d{\bf x}$ centered at the physical position ${\bf x}$ with a
physical velocity within the volume element $d{\bf v}$ centered at
${\bf v}$ at time $t$. In the mean field approximation, it obeys
Vlasov's equation (e.g., Binney \& Tremaine 1987) (summation over repeated
indices is understood):

\begin{equation}
\frac{\partial f}{\partial t} + v_{i} \frac{\partial f}{\partial x_{i}} + g_{i} \frac{\partial f}{\partial v_{i}} = 0 \;\;\; ,
\label{vlasov}
\end{equation}

\noindent
where the mean--field gravitational field strength ${\bf g}$ is
determined self--consistently by the equations (the comma denotes spatial
derivative, i.e., $_{,i} \equiv \frac{\partial}{\partial x_{i}}$):

\begin{gather}
\subnumbers 
g_{i,i} ({\bf x}, t) = \Lambda - 4 \pi G m N \int d{\bf v} f({\bf x},
 {\bf v}, t) \;\; ,
\label{poissona}\\\stepsubnumber
g_{i,j} ({\bf x}, t) = 
g_{j,i} ({\bf x}, t)\;\;; 
\label{poissonb}
\end{gather}
\nosubnumbers

\noindent
$G$ denotes the gravitational constant and $\Lambda$ the 
cosmological constant.

We now define the averaging operation $\langle ... \rangle$: If
$\phi({\bf v})$ is any function of the velocity ${\bf v}$, then

\begin{equation}
\langle \phi \rangle := 
{m N\over \varrho({\bf x} , t)} \int d {\bf v} \phi({\bf v}) f({\bf x}, 
{\bf v} ,t) \;\;\; ,
\label{averaging}
\end{equation}

\noindent 
where the mass density is $\varrho = m N \int d{\bf v} f$. We define the
average velocity ${\bf {\bar v}} := \langle {\bf v} \rangle$, and the {\it
stress tensor} ${\bf \Pi} = \varrho ( \langle {\bf v v} \rangle - {\bf
{\bar v}{\bar v}})$, which takes velocity dispersion into account, i.e., the
fact that at a given position there are several particles with different
velocities. (It should be noticed that ${\bf \Pi}$ is a
positive definite, symmetric tensor).

From Vlasov's equation (\ref{vlasov})
the equations obeyed by the density $\varrho$, the average velocity
${\bf {\bar v}}$ and the stress tensor ${\bf \Pi}$ are found by integrating
out (\ref{vlasov}), $v_j \cdot$(\ref{vlasov}), and 
$v_j v_k \cdot$(\ref{vlasov}) over velocity space:

\begin{gather}
\subnumbers
\partial_{t} \varrho + (\varrho {\bar v}_{i})_{,i} = 0 \;\;\; ,
\label{momenteqa}\\\stepsubnumber
\partial_{t} {\bar v}_{i} + {\bar v}_{j} {\bar v}_{i,j} = g_{i} - 
{1 \over \varrho} \Pi_{ij,j} \;\;\; ,
\label{momenteqb}\\\stepsubnumber
\partial_{t} \Pi_{ij} + {\bar v}_k \Pi_{ij,k} + {\bar v}_{k,k} \Pi_{ij} = 
- \Pi_{ik} {\bar v}_{j,k} - \Pi_{jk} {\bar v}_{i,k} - L_{ijk,k} \;\;\; ,
\label{momenteqc}
\end{gather}
\nosubnumbers

\noindent
where $L_{ijk} = \varrho \langle (v_{i}-{\bar v}_{i})(v_{j}-{\bar
v}_{j})(v_{k}-{\bar v}_{k}) \rangle$. Eq. (\ref{momenteqb})
is known as Jeans' equation in stellar systems theory. This is not a
closed set of equations for the density $\varrho$, the average
velocity ${\bf \bar v}$ and the stress tensor ${\bf \Pi}$ and
(together with (2)) the gravitational field strength ${\bf
g}$, because of $L_{ijk}$. We may in fact find the equation obeyed by
this third--rank tensor, but it turns out to depend on the fourth
moment of the velocity and so on. We thus get an infinite hierarchy of
equations for all moments of the velocity.

The task is now to find a way to close this hierarchy. In
hydrodynamics the common practice is to assume that collisions lead to
local equilibrium on a short time--scale, so that the hydrodynamical
system of equations is reduced to balance equations on a much larger
time--scale for the five collisional invariants $\varrho$, ${\bar
v}_i$ and the internal energy $U$. In dilute gases, the Chapmann--Enskog
expansion is then applied to find Navier--Stokes' equation and Fourier's
law (e.g., Balescu 1991). This method cannot be employed in our case,
because collisions are neglected and local equilibrium is thus
undefined.

The simplest way to close the hierarchy without resorting to local
equilibrium considerations is to neglect velocity dispersion
altogether, i.e., $f\propto \delta ({\bf v} - {\bf {\bar v}})$ which is 
the definition of ``dust matter''
(e.g., Gurevich \& Zybin 1995). Then
$\Pi_{ij} = 0$, $L_{ijk} = 0$ and Eqs. (4) reduce to

\begin{gather}
\subnumbers
\partial_{t} \varrho + (\varrho {\bar v}_{i})_{,i} = 0 \;\;\; ,
\label{momentdusta}\\\stepsubnumber
\partial_{t} {\bar v}_{i} + {\bar v}_{j} {\bar v}_{i,j} = g_{i} \;\;\; ,
\label{momentdustb}
\end{gather}
\nosubnumbers

\noindent
which is the well--known set of equations (together with
(2)) employed to model gravitational 
structure formation, being equivalent to following the motion of individual
particles.

As remarked in the introduction, these equations have the problem that
their solutions develop caustics, precisely because of neglecting
velocity dispersion. We go one step further by explicitly including 
velocity dispersion from the beginning. Let us therefore assume that
there is a small velocity dispersion of order $\varepsilon << 1$.
This means that the typical relative deviations of particle
velocities from the average velocity ${\bf \bar v}$ are of order
$\varepsilon$ (i.e., $|{\bf v}_{typ} - {{\bf \bar v}}|
\sim \varepsilon |{{\bf \bar v}}|$). This assumption implies that $\Pi_{ij} 
\sim \varrho\varepsilon^2 |{{\bf \bar v}}|^2$ and $L_{ijk} \sim 
\varrho\varepsilon^3 |{{\bf \bar v}}|^3$. To study the effect of 
velocity dispersion, we shall retain only the leading order terms due
to it, i.e. we shall work only up to order $\varepsilon^2$.
\footnote{It must be remarked that this is {\em not} equivalent to assuming 
that the velocities be Gaussian distributed. Gaussianity imposes an
additional constraint on the set of Eqs. (6).}
Then $L_{ijk,k}$ can be dropped in Eq. (\ref{momenteqc})
provided $L_{ijk}$ changes spatially on the same scale as ${\bar v}_i$ and we
get (together with (2)) 
a closed set of equations for
$\varrho$, ${\bf \bar v}$, ${\bf \Pi}$ and ${\bf g}$:

\begin{gather}
\subnumbers
\partial_{t} \varrho + (\varrho {\bar v}_{i})_{,i} = 0 \;\;\; ,
\label{momentapproxa}\\\stepsubnumber
\partial_{t} {\bar v}_{i} + {\bar v}_{j} {\bar v}_{i,j} = 
g_{i} - {1 \over \varrho} \Pi_{ij,j} \;\;\; ,
\label{momentapproxb}\\\stepsubnumber
\partial_{t} \Pi_{ij} + {\bar v}_k \Pi_{ij,k} + {\bar v}_{k,k} \Pi_{ij}
= - \Pi_{ik} {\bar v}_{j,k} - \Pi_{jk} {\bar v}_{i,k}  \;\;\; .
\label{momentapproxc}
\end{gather}
\nosubnumbers

\noindent
Using this set of equations we go one step further down the hierarchy
of equations for the velocity moments.  It is, however, still
difficult to cope with these equations and, in the present paper, we
introduce the strong simplification that the velocity dispersion is
approximately isotropic, i.e. the stress tensor $\Pi_{ij}$ is diagonal
and has only one independent component (a pressure--like term):

\begin{equation}
\Pi_{ij} \approx p \delta_{ij}\;\;\;\;,\;\;\;\; p \;>\; 0\;\;\;\;.
\label{pressure} 
\end{equation}

This assumption is certainly sensible at early stages of the structure
formation process: we can think of an initial condition with isotropic
velocity dispersion (it has to be so in the case of a
homogeneous--isotropic matter distribution). Then, at earlier stages
of the evolution, it will remain approximately isotropic. At later
stages we have to consider the isotropy of the stress tensor as an
{\it idealization} of the generally anisotropic velocity ellipsoid.
This assumption is to be considered a weak point in our analysis,
since there is nothing to prevent the velocity distribution from
becoming largely anisotropic in contrast to the situation in a
collisional system; it may work for single galaxy halos, but the bulk
of the medium is not isotropic\footnote{From the observational point of view
velocity dispersion can only be inferred from the line--of--sight
information, the isotropy assumption is then employed when
concluding on the 3D dispersion value.}.
Isotropy may, however, provide a good model if violent
relaxation is taken into account (see Lynden--Bell 1967,
Henriksen \& Widrow 1997, White 1996 and Kull et al. 1997).
 
\noindent
With this assumption, Eq. (\ref{momentapproxc}) is greatly simplified:

\begin{equation}
\left( \dot{p} + {5 \over 3} \bar{v}_{i,i} p \right)\delta_{ij} + 
2 p \sigma_{ij} 
\;=\;0\;\;\;,
\label{isotropy}
\end{equation}

\noindent
where we have introduced the total or Lagrangian derivative 
along integral curves of the {\it mean} velocity field, $\; \dot{ } :=
\partial_{t} + {\bar v}_{i} \partial_{i}$, and the shear tensor 
$\sigma_{ij}$ by splitting the symmetric part of the mean--velocity gradient: 
${\bar v}_{(i,j)} = {1\over 2}({\bar v}_{i,j} + {\bar v}_{j,i}) = 
{1 \over 3}{\bar v}_{k,k}\delta_{ij} + \sigma_{ij}$. 

Notice that Eq. (\ref{isotropy}) imposes a strong kinematical
restriction, namely, $\sigma_{ij}=0$. Although we generalize the
basic system of equations by allowing for a nonvanishing $p$, the
interpretation of $p$ in terms of a strictly isotropic velocity
dispersion will accordingly restrict the kinematics of the mean
motion. This makes clear that, unless we do not consider $p$ as a
phenomenological generalization of the ``dust matter model'', the
assumption (\ref{pressure}) is far too restrictive. It is, however,
one of the assumptions which {\it has to} be imposed in order to
recover the ``adhesion approximation''.

Keeping this in mind we go on by evaluating the trace    
of Eq. (\ref{isotropy}); it yields a relation between the density 
and the ``dynamical pressure'':

\begin{equation}
\dot{p} = \partial_{t} p + {\bar v}_{i} p_{,i} = - {5 \over 3} p 
{\bar v}_{i,i} \;\;\; .
\end{equation}

\noindent
Combining this equation with the continuity equation (\ref{momentapproxa}) 
we find

\begin{equation}
\dot{p} = {5 \over 3} {p \over \varrho} \dot{\varrho} \;\;\;.
\label{stateequation}
\end{equation}

\noindent
For this equation it is straightforward to find its general 
(Lagrangian) integral (which we may interprete as an `equation of state' for
a single volume element):

\begin{equation}
p ({\bf X},t) = \kappa ({\bf X}) \varrho ({\bf X},t)^{5/3}\;\;\;,
\label{integral}
\end{equation}

\noindent
where $\kappa$ is positive and of order $\varepsilon^2$, and ${\bf X}$
are Lagrangian coordinates which label fluid elements and coincide with
the Eulerian coordinates ${\bf x}$ at the initial time $t=t_0$.
(Note that the integration of (\ref{stateequation}) implies that 
$\kappa \ne 0$; $\kappa = 0$ is a singular limit of the equations under
consideration.) 

If the physical system were a true fluid, (\ref{integral}) would
be interpreted as describing the adiabatic evolution of each volume
element in an ideal gas (consistently with the fact that we have
assumed $L_{ijk,k}=0$, which would be interpreted as the absence of
heat flux between neighbouring volume elements). But unlike in a
fluid, there are no collisions in the system we are studying that
prevent velocity dispersion from growing and/or becoming highly
anisotropic.

In the sequel we restrict the general integral (\ref{integral}) to the
case where the initial data $p({\bf X},t_0)$ and $\varrho({\bf
X},t_0)$ are chosen such that $\kappa$ is independent of ${\bf
X}$. Therefore, we drop the freedom of giving $p({\bf X},t_0)$
independently of $\varrho({\bf X},t_0)$.  Thus, along any trajectory,
we have the relationship $p=\kappa\varrho^{5/3}$ with the same
constant $\kappa$.
Consequently, we have the same relation also at any point in Eulerian
space.  The integral then attains the status of a (global) `equation
of state' for the medium occupying Eulerian space.  
However, 
here, it is not imposed as an equation of state in
thermodynamics, but follows from the dynamical equations under
consideration and the assumptions employed.

Taking this result into Eq. (\ref{momentapproxb}) we get our final
set of equations:

\begin{gather}
\subnumbers
\partial_{t} \varrho + (\varrho {\bar v}_{i})_{,i} = 0 \;\;\; ,
\label{momentimpa}\\\stepsubnumber
\partial_{t} {\bar v}_{i} + {\bar v}_{j} {\bar v}_{i,j} = 
g_{i} - {5 \over 3}\kappa 
\varrho^{-1/3} \varrho_{,i} \;\;\; ,
\label{momentimpb}\\\stepsubnumber
g_{i,i} = \Lambda - 4 \pi G \varrho \;\;\;\;\;\; , 
\;\;\;\;\;\;\; g_{i,j} = g_{j,i} \;\;\; .
\label{momentimpc}
\end{gather}
\nosubnumbers

\noindent
The difference between the closed set of Eqs. (12)
and the one usually employed in cosmology (2,5) is the
pressure--like term $\propto \varrho^{-1/3} \varrho_{,i}$, which takes
into account (under the approximations stated above) velocity
dispersion (i.e., the dynamical reaction due to multi--stream
regions).

\section{Equation for the mean peculiar--velocity
relative to a background Hubble flow.}

We shall now show what novel features arise in the dynamical
evolution prescribed by (12) due to the
inclusion of the pressure--like term in contrast to the evolution
prescribed by (2,5).

In order to compare our model with the commonly studied approximation
schemes for large--scale structure we shall perform a change of
variables. We consider the homogeneous--isotropic solutions of the
basic equations (Friedmann--Lema\^\i tre backgrounds) characterized by
the expansion factor $a(t)$ (and Hubble's function $H=\dot{a} / a$),
and define `comoving coordinates' ${\bf q}$, an {\it average}
peculiar--velocity field ${\bf \bar u}$ and a (mean field)
gravitational peculiar--acceleration ${\bf w}$ as follows:

\begin{equation}  
{\bf q}: = {1 \over a} {\bf x} \; , \; {\bf \bar u}: =
{\bf \bar v} - H {\bf x} \; , \; {\bf w}: = {\bf g} +
{4 \pi G \varrho_H - \Lambda \over 3} {\bf x} \, ,
\end{equation}

\noindent 
where $\varrho_H(t)$ is the homogeneous background density (satisfying
$\dot{\varrho}_H + 3 H \varrho_H =0$), which coincides with the
spatially averaged density, if we impose periodic boundary conditions
on ${\bf \bar u}$ and ${\bf w}$; in this case the Hubble--flow exists
and is uniquely defined (see Buchert \& Ehlers 1997)\footnote{It is
straightforward to show that this result carries over to the case
where velocity dispersion is present, since the Hubble flow cancels in
the expression for ${\bf \Pi}$, ${\bf \Pi} = \varrho ( \langle {\bf v
v} \rangle - {\bf {\bar v}{\bar v}}) \equiv \varrho ( \langle {\bf u
u} \rangle - {\bf {\bar u}{\bar u}})$, i.e., ${\bf \Pi}$ is a tensor
field on the torus, and the divergence of the tensor appears in the
basic equations, i.e. the corresponding flux vanishes by integrating over
the torus.}.

In terms of these variables, Eqs. (12) become 
(spatial derivatives refer now to ${\bf q}$ and
time derivatives are taken at constant ${\bf q}$; hereafter, we drop
the bar above ${\bf u}$ for notational simplicity):

\begin{gather}
\subnumbers
\partial _{t} \varrho + 3 H \varrho + {1 \over a} (\varrho u_{i})_{,i} = 0
\;\;\; , 
\label{momentcomovinga}\\\stepsubnumber
\partial_{t} {u}_{i} + {1 \over a} {u}_{j} u_{i,j} + H {u}_{i} 
= w_{i} - {5\over 3}{\kappa\over a} \varrho^{-1/3} \varrho_{,i} \;\;\; ,
\label{momentcomovingb}\\\stepsubnumber
w_{i,i}=-4 \pi G a (\varrho - \varrho_H) \;\;\;\;\;\; , 
\;\;\;\;\;\;\; w_{i,j} = w_{j,i} \;\;\; .
\label{momentcomovingc}
\end{gather}
\nosubnumbers

\noindent
Combining (\ref{momentcomovingb}) and (\ref{momentcomovingc}) we find the 
following equation, using the Lagrangian derivative operator $\; \dot{ } :=
\partial_{t}\vert_x + {\bar v}_{i} \partial_{i}\vert_x = \partial_{t}\vert_q 
+ {u_i\over a} \partial_{i}\vert_q$ (written in vector form):

\begin{equation}
\dot {\bf u} + H {\bf u} = {\bf w} + \zeta \Delta {\bf w} \;\;\;, 
\label{mastereq1}
\end{equation}

\noindent
with the coefficient $\zeta = {5\over 3}{\kappa\over 4\pi G
a^2}\varrho^{-1/3} > 0$, which depends on density and explicitly on
time. The difference between Eq. (\ref{mastereq1})
and the one generally used to model large--scale structure formation
(which is found by setting $\zeta=0$) is the $\Delta {\bf w}$ term.

Recall first the case of ``dust'' (i.e., no velocity dispersion, $\zeta = 0$).
In the weakly nonlinear regime Zel'dovich's approximation (Zel'dovich 1970, 
1973) is a successful model until shell--crossing singularities develop.
The trajectories in that approximation obey the parallelism of 
peculiar--gravitational acceleration and peculiar--velocity (see, e.g. 
Bildhauer \& Buchert 1991, Kofman 1991, Buchert 1992),

\begin{equation}
{\bf w} = F(t) {\bf u} \;\;\;, \;\;\;F(t) = 
4\pi G\varrho_H {b(t)\over {\dot b}(t)} \;\;\;,
\label{parallelity}
\end{equation}

\noindent
where $b(t)$ is the growing mode solution of the linear theory of
gravitational instability for ``dust'' (i.e., it solves the equation
$\ddot{b} + 2 H \dot{b} - 4 \pi G \varrho_{H} b = 0$).

\noindent
Since Eqs. (14) were obtained under the
condition of small velocity dispersion, we can try to extrapolate
Zel'dovich's approximation (\ref{parallelity}) into this regime 
(which is equivalent to solving Eq. (\ref{mastereq1}) by iteration) and
thus find from Eq. (\ref{mastereq1}):

\begin{equation}
\dot {\bf u} + (H-F) {\bf u} = \zeta F(t) \Delta {\bf u} \;\;\;.   
\label{mastereq2}
\end{equation}

\noindent
In order to construct the model we have to
derive from the solution of Eq. (\ref{mastereq2}) the trajectory
field of the {\it average flow} ${\bf q} = {\bf F}({\bf X},t)$ by
quadrature: ${\bf u}=a {\dot{\bf F}}$. 
Changing the temporal variable from $t$ to $b$ (this is possible since
$b(t)$ is a monotonically increasing function of time) and defining a
rescaled velocity field $\tilde{\bf u} = {\bf u} / a
\dot{b}$, Eq. (\ref{mastereq2}) becomes

\begin{equation}
\frac{d \tilde{\bf u}}{d b} = {\mu} \Delta \tilde{\bf u} \;\;\; , 
\;\;\; (\; {d \over db} := {\partial \over \partial b} + 
\tilde {\bf u} \cdot \nabla \;) \;\;\; ,
\label{mastereq3}
\end{equation}

\noindent
where $\mu = \zeta F(t) / {\dot b} > 0$. If $\mu$ were independent of
density, this equation would become 3D Burgers equation, whose
solution is analytically known. The fact that $\mu$ depends on density
presents an obstacle for finding an analytical solution. The principal
advantage of Eq. (\ref{mastereq3}) over the same equation with
$\mu = 0$ is that it does not lead to caustic formation, since
velocity dispersion smoothes out the singularity (Zel'dovich \&
Shandarin 1982, Shandarin \& Zel'dovich 1989; see Ginanneschi 1998 and
ref. therein for a thorough analysis). Therefore, this equation could
be used as it stands to follow the dynamical evolution beyond the time
when singularities in the ``dust'' continuum would arise. To be more
precise, ``shell--crossing'' would still happen, but this doesn't
manifest itself as a singularity in the average flow, rather as a (smooth)
peak in the density field.
Morphologically distinct patterns, classified by the
Lagrange--singularity theory in the case of ``dust'' (see Arnol'd et
al. 1982), will also emerge in the sense of smoothed--out images of
critical sets on the Lagrangian manifold of the ``dust'' medium.
This picture might not be true for large velocity dispersion.
 
As time goes by and the system becomes more and more virialized,
Eqs. (14) cease to be a good description of the
dynamical evolution, because velocity dispersion generically both grows
and becomes anisotropic. 

Eq. (\ref{mastereq3}) is formally equivalent to the key equation of
the ``adhesion model'' in which the coefficient $\mu$ is positive and constant
in space and time 
(Gurbatov et al. 1989), and we see that the $\Delta {\bf u}$ term behaves as a
viscous force. In fact, in the singular limit $\mu \rightarrow
0$ (the so--called inviscid limit), we approach a singular case of 
the ``adhesion model''
(Gurbatov et al. 1989; see also Gurbatov et al. 1983, 1985), 
the so--called ``sticky particle model'' for
which geometrical construction techniques have been advanced (Pogosyan
1989, Kofman et al. 1990, Sahni \& Coles 1995). 
For further details on models based on Burgers' equation the reader 
may also consult the book by Gurbatov et al. (1991) and the interesting 
article by Vergassola et al. (1994). Further insight into possible 
applications is provided by a model of Jones (1996) for a two--component 
collisionless--baryonic system, which is related
to the ``adhesion model''. 

Although we here recover a variant of the ``adhesion model'', 
it should be clear that the new model presented implies an improvement
for various reasons. The main one is that the dissipative--like term
appears in a natural way together with a clear physical interpretation
(i.e., velocity dispersion), unlike in the ``adhesion model'', where
it is motivated by phenomenological arguments. 

This fact also solves the `momentum--conservation violation problem'
arising in the ``adhesion model'' (Kofman \& Shandarin 1988; see also
Gurbatov et al. 1983, Shandarin \& Sathyaprakash 1996).  Indeed, if
velocity dispersion is neglected, Eq. (\ref{momentdustb})
describes the evolution of the velocity of {\em individual} particles,
so that appending a $\Delta {\bf u}$ force to this equation implies
that momentum is not conserved. On the other hand, the velocity field 
${\bf u} ({\bf q}, t)$
is the {\em mean} velocity of the particles at the (comoving) Eulerian
position ${\bf q}$ at time $t$. For the mean velocity, gravity is not
the only force, there is also an effective force due to velocity
dispersion, so that the momentum balance is not violated. This
effective force (pressure) simply provokes a flow of kinetic energy
between the bulk average motion and the ``random'' motion of particles
(described by velocity dispersion).
The total kinetic energy (the kinetic energy of the mean flow plus
internal kinetic energy) changes only due to gravitational work.

The conservation of mass and momentum is already evident from the way we have 
found Eqs. (\ref{momentcomovinga}) and (\ref{momentcomovingb}), which 
are balance equations expressing the conservation of mass and momentum. 
In fact, since the pressure force appears as a gradient (in general, as
the divergence of a tensor for anisotropic velocity dispersion), it
describes the transfer of momentum between neighbouring volume elements.

The {\it reversibility} of the transfer of momentum can also be
immediately appreciated from our model equations. The coefficient
$\mu$ {\it is not related to an irreversible process}: if one performs
a time reversion $t \rightarrow -t$, ${\bf u} \rightarrow -{\bf u}$
(${\bf w}$, $\varrho$ remain unchanged), then $F \rightarrow -F$,
$\zeta \rightarrow \zeta$, and
Eq. (\ref{mastereq2}) is invariant under time reversion. (This is
true also for the exact equation (\ref{mastereq1}) as well as for the
basic system of equations. Therefore, the coefficient $\zeta$ should
not be considered as a viscosity, a true viscosity coefficient would
not change sign under time--reversion. Another difference is that the
coefficient $\mu$ depends on the initial conditions (through
$\kappa$), whereas true damping rates are insensitive to initial
conditions. We may call the coefficient $\mu$ {\it gravitational
multi--stream} or {\it GM--}coefficient, because it arises from the
self--gravitation of multi--stream systems. The phenomenon of
``reversible damping'' is not unknown in physics (e.g. Landau
damping); the phenomenon we are discussing coincides with
`non--dissipative gravitational turbulence' as described by Gurevich
\& Zybin (1995).

Another interesting point is that the GM--coefficient $\mu$ in
(\ref{mastereq3}) is not arbitrary: it has a dependence on time and
density which is determined by the dynamical equations.  However, this
(very plausible) fact complicates the model which no longer can be
solved by the known solution of the 3D Burgers equation.  We have not
found an immediate variable change to perform Cole--Hopf--type or
other transformations in order to reduce the model to the linear
diffusion equation. Application of the methods which work in the case
of the 3D Burgers equation to more general problems apparently
creates obstacles (see, e.g., Nerney et al. 1996).  The model must be
solved numerically by determining $\mu(\varrho,t)$ at each
time--step for each particle.

\section{Discussion and criticism.}

We have presented a system of equations that allows
studying the structure formation process at stages where the usually
employed Euler--Poisson system breaks down. We discussed the
possibility of constructing models for large--scale structure which do
not suffer from the occurrence of shell--crossing singularities.  We
also proposed a model which features ``adhesion'' of fluid elements
similar to the commonly used ``adhesion approximation'', and allows
following the structure formation process beyond stages where a
laminar fluid approximation based on ``dust'' breaks down
due to the development of multi--streaming: a ``viscous'' term appears
in the model equation which we derived from the combined action
of multi--stream flow and self--gravity.  

Besides the obvious advantages that we do not have to {\em invent} an
adhesive term and that momentum conservation is not violated, we also
see a further merit of this approach: we have gained some physical
insight into the ``adhesion model'' and could even improve on it.

One may ask why the combined effect of self--gravity and velocity dispersion
actually ``looks like'' a dissipative term. To clarify this, let us add the
following illustration:
consider a vessel, filled with a non--viscous toy fluid, and placed into the 
gravitational field of the Earth. Initially, the fluid may be at rest and its
density homogeneous. This state is clearly unstable and the fluid will evolve
into a stable, inhomogeneous and stratified state. The overall kinetic energy 
vanishes in this example, but the gravitational potential energy has decreased.
The ``missing energy'' has not been dissipated away, rather it has been 
transformed into internal kinetic energy due to compression work; the final 
volume of the fluid is smaller than the initial volume, and the 
pressure has always opposed the motion induced by gravity. This is similar to 
what happens in the cosmological setting: volume elements would have equal 
tendency to compress or expand due to pressure--like forces, but the presence
of gravity favors contraction of volume elements, especially when they are 
falling onto high--density regions. The effect of transformation of energy into
internal velocity dispersion is large, because not only potential energy, but
also bulk kinetic energy decreases, when the fluid elements crash onto
high--density walls.

We have also shown that pressure--like forces may be relevant even on
cosmological scales and an effective Jeans' length may be larger than
the (comoving) globular cluster scale, which is commonly attributed to
a hydrodynamical (ideal gas) pressure (compare this to the
effect of a gas pressure within pancakes investigated by Sunyaev \&
Zel'dovich 1972). However, we now understand clearly the limited
status of this and similar models. We shall now elaborate on this
criticism in more detail.

The extrapolation of the relationship (\ref{parallelity}) into the
weakly non--linear regime is well justified for ``dust''. If we define
the weakly non--linear regime as suggested in (Buchert 1989), namely,
as a linearization in Lagrangian space, Zel'dovich's extrapolation of
the Eulerian linear perturbation solution into the nonlinear regime is
not only consistent with Lagrangian linearization of the full
Euler--Poisson system (Buchert 1989, see also Doroshkevich et al. 1973
for a self--consistency test for part of the
system), but arises naturally as a subcase of first--order solutions
in a Lagrangian perturbation approach (Buchert 1992).  This
self--consistency of the ``dust model'' is mirrored in the fact that the
trajectories of Lagrangian first--order perturbations even provide a
class of 3D exact solutions to the Euler--Poisson system (Buchert
1989).

If pressure is taken into account, extrapolation of the ``dust'' model
trajectories (as we had to do in order to recover the ``adhesion
approximation'') is no longer permitted. As we have shown it may be
used as a {\it first} approximation for small velocity dispersion (and
this is precisely the method followed in Zel'dovich \& Shandarin
1982 and Bharadwaj 1996).  
A similar remark applies to other assumptions of extrapolation
like the Frozen--Potential--Approximation (Brainerd et al. 1993,
Bagla \& Padmanabhan 1994), which we may employ instead of
(\ref{parallelity}). This approach simply follows the evolution of
velocity dispersion along the trajectories dictated by the ``dust''
model, its main drawback being that the ``back--reaction'' of velocity
dispersion on these trajectories is not taken into account. In fact,
the simple version of parallelism (\ref{parallelity}) 
is not even justified in the (Eulerian) linear regime
(Buchert et al. 1998). The break--down of the assumption
(\ref{parallelity}) after pancake formation is plausible and it is
well--known for epochs after shell--crossing to which we want to apply
the model (Doroshkevich 1973).

\noindent
Nevertheless, the ``adhesion approximation'' works rather
well, because the model is taken at face value and not restricted {\it
a posteriori} by the constraints which led us to its
derivation. Accordingly, we expect the model we proposed to improve on
the ``adhesion approximation'', since it specifies the dependence of
``adhesion'' on the local density.  The publication of this work was
motivated by this gain of insight and improvement.
Let us summarize the constraining assumptions which restrict the general
problem to the ``adhesion approximation'':

\smallskip\noindent$\bullet\;$
small velocity dispersion; we keep only the first 
equation in the hierarchy of velocity moments (A1).

\noindent$\bullet\;$
isotropy of the dispersion tensor which implies isotropy of the
mean motion (A2).

\noindent$\bullet\;$
parallelity of peculiar--velocity and --acceleration, both defined 
relative to a global Hubble--flow (A3).

\noindent$\bullet\;$
a spatially constant relationship between initial density and 
initial pressure (A4).

\noindent$\bullet\;$
the GM--coefficient is constant in space and time as well as positive
(A5) (which could be achieved by {\em imposing} a relationship $p
= \kappa \varrho^2$).

\smallskip
\noindent
This list of assumptions shows that there is enough room for a
generalization of the ``adhesion approximation''. 

\noindent 
Our main concern should be focussed on assumptions (A1) and (A2). Although
both of them may be considered a good working hypothesis from a theoretical
point of view, we know that the physical situation is not in favor of these
assumptions, especially if we follow the average flow further into the
nonlinear regime. Velocity dispersion will become quickly large in the
multi--stream regime. A hint of this can be seen in observations of velocity
dispersion in rich clusters of galaxies which is of the order of 1000 km/s, 
i.e. certainly not small compared with the bulk speed. However, viewing
assumptions (A1) and (A2) together, we would obtain substantially similar
models by 
arguing on phenomenological grounds: assumption (A2) gives an idealized model 
for the dispersion tensor and it just needs a relationship $p = \beta
(\varrho)$ to close the hierarchy of velocity moments. In light of a 
phenomenological approach assumption (A1) or, alternatively, assumption (A5)
just specify the function $\beta$,
but we may as well consider other relationships as discussed in (Buchert et al.
1998). This problem calls for a better model that relates the dispersion 
tensor to the density and velocity fields, but it does not invalidate the
main ideas exemplified with assumption (A1).

Concerning assumption (A3) a first indication of how to
proceed in order to investigate a valid extrapolation into the weakly
nonlinear regime is furnished by Eq. (\ref{mastereq1})
itself: the adhesive term is proportional to $\Delta {\bf w}$ rather
than $\Delta {\bf u}$. It is possible to derive an evolution equation
for the peculiar--gravitational field strength ${\bf w}$. Its
specification to the weakly nonlinear regime, however, requires
Lagrangian perturbation techniques that lie beyond the scope of the
present work. Assumptions (A4), (A5) have their origin in
mathematical simplicity and could be easily relaxed in a numerical
simulation of the model.

\begin{acknowledgements}

The authors wish to thank Juan P\'erez--Mercader for enlightening
discussions, Claus Beisbart, J\"urgen Ehlers, Bernard Jones, Martin Kerscher,
Lev Kofman, Adrian Melott, Sergei Shandarin, Mikel Susperregi, David Weinberg,
Kirill Zybin and the referee for valuable comments. 

\noindent
TB is supported by the
``Sonderforschungsbereich SFB 375 f\"ur Astro--Teilchenphysik''
and thanks LAEFF for hospitality
and generous support during working visits in Madrid. AD 
thanks ``SFB 375'' for support during working visits in Munich.

\end{acknowledgements}

\end{document}